\documentclass{article}
\usepackage{amsmath,graphicx,subfig}
\usepackage{caption}
\usepackage{amssymb,amsbsy}
\usepackage{tikz}
\author{P.J.Arias, Adel Khoudeir and J. Stephany}
\title{Unitary spin 2 models  in 3-D}

\begin{document}
\tikzstyle{bag} = [text width=2em, text centered]
\tikzstyle{bag1} = [text width=5em, text centered]
\tikzstyle{end} = []

\begin{titlepage}

\hfill{ \parbox{3cm}{{AEI-2012-003}\newline {SB/F/389-11}}} 
\hrule \vskip 2cm

\begin{center}{\large{\bf Master actions for linearized massive \\ gravity  models  in 3-D }}
\vskip 0.5cm
Pio.J.Arias$^a$, Adel Khoudeir$^b$ and J. Stephany$^{c,d}$
\\ \vskip 0.2cm
{\it $^a$Centro de F\'{\i}sica Te\'orica y Computacional, Facultad de Ciencias, \\
Universidad Central de Venezuela, AP 20513,Caracas 1020-A, Venezuela}
\\ \vskip 0.3cm
{\it $^b$Centro de F\'{\i}sica Fundamental, Departamento de F\'{\i}sica, \\
Facultad de Ciencias, Universidad de Los Andes, M\'erida 5101, Venezuela}
\\ \vskip 0.3cm
{\it $^c$Max-Planck-Institut f\"ur Gravitationsphysik,Albert-Einstein-Institut\\
Am M\"uhlenberg 1, 14476 Golm, Germany}\\
\vskip 0.3cm
{\it $^d$Departamento de F\'{\i}sica, Universidad Sim\'{o}n Bol\'{\i}var, \\
Apartado Postal 89000, Caracas 1080-A, Venezuela}\\
\vskip 0.3cm
{\it e-mail: pio.arias@ciens.ucv.ve, adel@ula.ve, stephany@usb.ve}
\end{center}
\vskip 1cm
\begin{abstract}
We present a unified analysis of   the self-dual, second order, topologically massive and the recently introduced fourth order models of massive gravity  in $3D$. We show that there is a family of first order actions  which interpolate between these different single excitation models. We show how the master actions are related by duality transformation. We construct by the same method the master action which relates the fourth order  new massive model with two excitations and the usual second order model with Fierz-Pauli mass. We show that the more general model obtained by adding a Chern-Simons term to the new massive model is equivalent off shell to the second order spontaneously broken linearized massive gravity. 
\end{abstract}
 \hskip 1cm
\begin{center}
Keywords:{\it Massive gravity,self duality,topological mass}\\
\noindent PACS numbers: 11.10 Kk, 04.60.Kz, 04.60.Rt
\end{center}
\vskip 1cm
\hrule
\bigskip
\vfill
\end{titlepage}

\section{Introduction}

Three dimensional massive tensor models have a rich structure which  emerged in the three decades since the discovery  of the topologically massive model (TMM) \cite{DesSJT1982}.  This model, in which the non propagating Einstein-Hilbert action is complemented with a gravitational Chern Simons term, yields  upon linearization a third order  gauge invariant model which  propagates a single parity sensible excitation. Other unitary models with a single excitation were then recognized. The  self-dual model (SDM),    which  upgrades to the case of spin 2,  the ideas used to construct the vectorial self dual theory \cite{TowPPV1984}  equivalent to  topologically massive electrodynamics \cite{DesSJ1984} was introduced in Ref. \cite{AraCK1986}. It provides a first order equation for the  tensor  field . A second order model with the same spectrum usually denominated the intermediate model (IM) was also formulated in the same work. Master actions related the SDM to the IM and the IM to the TMM were presented. The IM allows for a curved space \cite{AraCK1994} formulation which consists of the ordinary Einstein Hilbert action and a first order, vector like quasi topological term constructed out of the dreibein variables. This model is diffeomorphism invariant but has the local Lorentz invariance  spontaneously broken \cite{AraCAK1997}. More recently a  Lorentz, gauge and conformal fourth order model (FOM)   which also describes a single massive mode has been presented \cite{DalDM2009}. This model is not unrelated  with the also recently introduced and very interesting system  dubbed new massive gravity (NMG)\cite{BerEHT2009} \cite{BerEHT2009b}  which describes in an elegant and gauge invariant form the two physical excitations of a massive, parity invariant spin 2 particle. These theories are ghost free ,  despite of the fact that their field equations contain higher terms in the curvature in contrast to equivalent situation in  $D\geq 4$ \cite{SteK1977} \cite{SteK1977b}). This has motivated to review \cite{DesS2009} in this context the standard relation between negative energy excitation and higher order equations as was already proposed elsewhere \cite{BenCM2008}.

The TMM and the NMG  have in common that the Einstein-Hilbert term  appears with a sign  opposite to the conventional one in $D \geq 4$ gravity. Combining these two theories,  the most general unitary action for massive spin 2 \cite{AndRBR2010} is constructed and shown to describe two physical excitations with different masses  in a gauge invariant way. From this theory, appears in a particular limit the FOM \cite{DalDM2009}, \cite{AndRBR2010}. All these models may be formulated for generic gravitational background and when the background is AdS, new interestingly  phenomena arise \cite{LSS}. The linearization on flat and AdS backgrounds have been studied in references \cite{BerEHT2009b} and \cite{DesS2009}.

As already mentioned, in a form  which is analogous to the case of vector fields where the  self dual theory \cite{TowPPV1984} is equivalent to  topologically  massive  electrodynamics \cite{DesSJ1984},  at linearized level the self dual massive spin 2 \cite{AraCK1986}  is equivalent to the topological massive gravity. The self dual model do not have any local invariance and their action is first order. Nevertheless, it may be viewed as gauge fixed formulation of their related higher order models\cite{GiaRRS1991,RestAS1993,AriPS1995}. Classical and quantum aspects of these dual equivalences have been studied  in references \cite{DalDM2009b,DalD2009,DalDM2010,AriasF2011}.

In section 2 we take a systematic approach to establish the equivalences, at the linearized level in topological trivial manifolds, of the models mentioned both for one and two physical excitations. For the family of parity sensible models we display  a network of master actions ultimately relating the SDM action to the FOM. The process of increasing the order of the action is accomplished with a corresponding improvement of the symmetries in the chain of master actions.  Although the models obtained from the linearization of metric curved gravity models may be written in terms of symmetric tensor fields the structure discussed in this article is best shown using non symmetric tensors. This is evident in the discussion of section 3 where the master actions are shown to emerge from the application of dual transformations to the different mode.   In section 4 we construct using the duality transformation, the master action which forces the equivalence of the NMG with the usual massive Fierz Pauli model discussed in \cite{BerEHT2009}. We also show the  equivalence of the most general fourth order model \cite{AndRBR2010}, \cite{DalDM2009c}  with the second order model considered in ref \cite{AraCAK1997} which propagates unitarily two different masses.  Since all the actions presented here are quadratic in the fields, the discussion of these equivalences in the framework of path integral functional  may also be pursued.

\section{Models with one excitation}

We start with the self-dual (SDM) first order action action \cite{AraCK1986}
\begin{equation}\label{I0}
    I_0[w]=-\frac{m}{2}<w_{\mu\nu} w^{\nu\mu}-w_\mu^{\ \ \mu} w_\nu^{\ \ \nu}>+\frac{1}{2}<w_{\mu\lambda} \epsilon^{\mu\nu\rho}\partial_\nu w_\rho^{\ \ \lambda}>,
\end{equation}
where $<>$ denotes integration in $3D$.
This model is equivalent to the second order intermediate model (IM) which can be written in the two alternative forms,
\begin{eqnarray}\label{In}
  I^{2th} [h]&=& \frac{1}{2m}<\epsilon^{\nu\alpha\beta}\partial_\alpha h_{\beta\rho} \epsilon^{\rho\mu\sigma}\partial_\mu h_{\sigma\nu}>-\frac{1}{4m}<(\epsilon^{\alpha\beta\gamma}\partial_\beta h_{\gamma\alpha})^2>\nonumber\\
&-&\frac{1}{2}<h_{\mu\lambda} \epsilon^{\mu\nu\rho}\partial_\nu h_\rho^{\ \ \lambda}>\\ \nonumber
&=& \frac{1}{2m}<h_{\nu}^{\ \rho} \epsilon^{\nu\alpha\beta}\partial_\alpha W_{\beta\rho}[h] >
-\frac{1}{2}<h_{\mu\lambda} \epsilon^{\mu\nu\rho}\partial_\nu h_\rho^{\ \lambda}>\nonumber \ .
\end{eqnarray}
with 
\begin{eqnarray}
W_{\mu \nu}[h] &\equiv& (W_{\mu \nu})^{\rho \sigma} h_{\rho \sigma}\nonumber\\
&= &[\delta_\nu ^\alpha \delta _\mu ^\sigma - \frac{1}{2}\eta_{\mu\nu} \eta^{\alpha\sigma}]  \epsilon_ \alpha ^{\ \tau\rho}\partial_\tau h_{\rho \sigma}\ \ .
\end{eqnarray}
This is the linearized version of the massive vector Chern-Simons  gravity \cite{AraCK1994} whose action is given by the Einstein-Hilbert action and a vectorial Chern-Simons term constructed out of the dreibein variables. The IM is invariant under gauge transformations, $\delta h_{\mu\nu} = \partial_\mu \zeta_\nu$. Equivalence with the SDM 
may  be shown in different ways. For example a detailed analysis of the constraints structure of the models lead to interpret the self-dual model as a gauge fixed formulation of the intermediate model \cite{AriPS1995}. These models may be also related by a duality transformation \cite{DalDM2009b,AriasF2011} which establishes the equivalence on topologically trivial manifolds  leaving room to a more subtle relation on general manifolds as it happens with the vector models
\cite{AriPR1996,AriPRL1996,SteJ1997}. The dual relation between the models is signalized by the existence of a master action which provides the most direct demonstration of their equivalence. This master action, which is the analogous of the one used by Deser and Jackiw in their treatment of the vector models \cite{DesSJ1984}  was given in \cite{AraCK1986} 
\begin{eqnarray}\label{I1}
   I_1 [w,h]&=&-\frac{m}{2}<w_{\mu\nu} w^{\nu\mu}-w_\mu^\mu w_\nu^\nu>+<w_{\mu\lambda} \epsilon^{\mu\nu\rho}\partial_\nu h_\rho^{\ \lambda}>\nonumber\\&-&\frac{1}{2}<h_{\mu\lambda} \epsilon^{\mu\nu\rho}\partial_\nu h_\rho^{\ \lambda}>.
\end{eqnarray}
in terms of general tensor fields $h_{\mu\nu}$ and $w_{\mu\nu}$  neither symmetric or antisymmetric. Note that the  structure of this action, is given by a Fierz-Pauli mass term for one of the fields plus a term involving the rotor of both fields ending with a vector Chern Simons like term. This structure is exploited below to generalize this master action. Either the SDM or the IM may be derived covariantly by substituting the equations of motions but this procedure should be supplemented by a more careful analysis to guarantee the canonical equivalence of the systems\cite{AraCK1986,AriPS1995}. To obtain the IM one takes variations in (\ref{I1}) respect to $w$ and obtains the identity
\begin{equation}\label{h(w)}
 w_{\rho\mu}(h)=   \frac{1}{m}W_{\mu \nu}[h] \ \ .
\end{equation}
This equation may be viewed as a kind of self dual change of variables and will be used repeatedly in what follows.
Substituting (\ref{h(w)})  in (\ref{I1}) leads to the second order intermediate model (\ref{In}).
To recover the SDM one considers the second of the equations of motions of (\ref{I1}),
\begin{equation}\label{Rotor}
 \epsilon^{\mu\nu\rho}\partial_\nu w_{\rho\sigma}= \epsilon^{\mu\nu\rho}\partial_\nu h_{\rho\sigma}\ \ .
\end{equation}
Substitution of this in $I_1$ gives the SDM. Why this covariant treatment works may be understood by noting that (\ref{Rotor})  forces the transverse components of $w_{\mu\nu}$ and $h_{\mu\nu}$ to be equal and then,
\begin{equation}\label{h=wl}
w_{\mu\nu}- h_{\mu\nu} = \partial_\mu \lambda_\nu\ \ .
\end{equation}
Setting  $\lambda_\nu = 0$ using the gauge invariance of $I_1$, 
\begin{equation}
 \delta h_{\mu\nu} = \partial_\mu \zeta_\nu; \quad  \delta w_{\mu\nu} = 0,
\end{equation}
it follows that
\begin{equation}\label{h=w}
w_{\mu\nu} = h_{\mu\nu} .
\end{equation}
Combined with (\ref{h(w)}) this implies
\begin{equation}\label{hW}
    h_{\mu \rho} =  \frac{1}{m}[\epsilon^{\rho\tau\sigma}\partial_\tau h_{\sigma\mu}-\frac{1}{2}\eta_{\rho\mu}\epsilon^{\lambda\tau\sigma}\partial_\tau h_{\sigma\lambda}] = \frac{1}{m}W_{\mu \rho}[h]\ \ .
\end{equation}
These are the equations of motion of the self dual theory. Alternatively one may note that $I_1$ does not depend on the longitudinal part of $h_{\mu\nu}$.

We can iterate the mechanism which forces the equivalence between $I_1$ and $I_0$ and introduce a second and a third master actions $I_2$ and $I_3$  given respectively by 
\begin{eqnarray}\label{I2}
  I_2 [w,h,v]&=& -\frac{m}{2}<w_{\mu\nu} w^{\nu\mu}-w_\mu^\mu w_\nu^\nu>+<w_{\mu\lambda} \epsilon^{\mu\nu\rho}\partial_\nu h_\rho^{\ \lambda}>\nonumber\\
&-&<h_{\mu\lambda} \epsilon^{\mu\nu\rho}\partial_\nu v_\rho^{\lambda}> +\frac{1}{2}<v_{\mu\lambda} \epsilon^{\mu\nu\rho}\partial_\nu v_\rho^{\ \lambda}>.
\end{eqnarray}
\begin{eqnarray}\label{I3}
  I_3 [w,h,v,u]&=& -\frac{m}{2}<w_{\mu\nu} w^{\nu\mu}-w_\mu^\mu w_\nu^\nu>+<w_{\mu\lambda} \epsilon^{\mu\nu\rho}\partial_\nu h_\rho^{\ \lambda}>\nonumber\\
&-&<h_{\mu\lambda} \epsilon^{\mu\nu\rho}\partial_\nu v_\rho^{\lambda}>+<v_{\mu\lambda} \epsilon^{\mu\nu\rho}\partial_\nu v_\rho^{\lambda}>\\ &-&\frac{1}{2}<u_{\mu\lambda} \epsilon^{\mu\nu\rho}\partial_\nu u_\rho^{\ \lambda}\nonumber>.
\end{eqnarray} 
In each case the vector Chern Simons term is re-expressed with apparent redundancy  using an auxiliary field. One may also view the procedure as a duality transformation as will be discussed below. 
The relation between $v_{\mu\nu}$ and $h_{\mu\nu}$  in $I_2$ have the same structure that the one between $h_{\mu\nu}$ and $w_{\mu\nu}$ in $I_1$ just discussed. Following the same steps one obtains that $I_2[w,h,w(h)]=I_1[w,h]$ with $w(h)$  defined by (\ref{Rotor}). This means that $I_2$ is  equivalent to $I_1$ and forcefully to $I_0$ and $I^{2th}$.
Analogously it is straightforward to show that $I_3[w,h,w,u(w)]=I_2[w,h,u]$ with $u(w)$ defined now by the equation of motion with the same structure that (\ref{Rotor}) and its corresponding equivalence with all the previous actions. Below  we show how these master actions may be used to discuss the equivalence of all the single excitation models of linearized gravity in 3D. Other actions $I_N$, written  in terms of more fields may also be constructed by the same mechanism and are all equivalent, but when reduced  to models with a single field they do not lead to new unitary models different of the four discussed in this paper. In particular, the models of order higher than four in the derivatives which may be associated  with them using covariant methods, propagate ghost excitations.

To continue our discussion let us show how $I_2$ may be related to third order TMM  whose action is given by \cite{DesSJT1982} 
\begin{equation}\label{v}
I^{TMM}[h] = -\frac{1}{2m} <h_{\mu\lambda} \epsilon^{\mu\nu\rho}\partial_\nu W_{\rho}^{\ \lambda}[h]>+\frac{1}{2m^2}<W_{\mu\lambda}[h] \epsilon^{\mu\nu\rho}\partial_\nu W_{\rho}^{\ \lambda}[h]>.
\end{equation}
Here the first term is the linearized Einstein-Hilbert action with the opposite sign and the second term is the linearized true Chern-Simons gravitational action. This  action can also  be expressed alternatively in terms only of the symmetric part $H_{\mu\nu}=\frac{1}{2}(h_{\mu\nu}+h_{\nu\mu})$ \footnote{We use also the Einstein tensor $G_{\mu\nu}= \epsilon_\mu ^{\rho\sigma}\partial_\rho W_{\sigma\nu}[h] = \epsilon_\mu ^{\rho\sigma} \epsilon_\nu ^{\alpha\beta}\partial_{\rho}\partial_{\alpha}H_{\sigma\beta}$, so that $G_{\mu\nu}=\Box H_{\mu\nu} + ...$. The Schouten tensor is $S_{\mu \nu} = (W_{\mu \nu})^{\rho \sigma} W_{\rho \sigma}[h]$ and the Cotton tensor, which is symmetric, transverse and traceless is defined by  $C^{\mu}_{\nu} \equiv \epsilon^{\mu\rho\sigma}\partial_\rho S_{\sigma\nu}$, with $S_{\mu\nu} \equiv G_{\mu\nu} - \frac{1}{2}\eta_{\mu\nu} G = R_{\mu\nu} - \frac{1}{4}\eta_{\mu\nu}R$  }. The TMM may be obtained directly from the SDM by performing the the self-dual change of variables $w(h)$ of equation (\ref{h(w)}) in (\ref{I0}). The connection with $I_2$ goes through the following action 
\begin{eqnarray}\label{In2}
   I^{2th}_2 [h,v]&=& \frac{1}{2m}<h_{\nu \rho} \epsilon^{\nu\alpha\beta}\partial_\alpha W_{\beta\rho}[h] >\nonumber\\
&-&<h_{\mu\lambda} \epsilon^{\mu\nu\rho}\partial_\nu v_\rho^{\ \lambda}>+\frac{1}{2}<v_{\mu\lambda} \epsilon^{\mu\nu\rho}\partial_\nu v_\rho^{\ \lambda}>
\end{eqnarray}
This is obtained  substituting  (\ref{h(w)}), which is again one of the equations of motion, in (\ref{I2}).This procedure is the same used to obtain the second order  model (\ref{In})  from $I_1$. $I^{2th}_2 [h,v]$   works  as a master action for $ I^{2th}$ and $ I^{TMM}$ and is equivalent to the slightly different one introduced in \cite{AraCK1986}. One of the equations of motion states that the transverse parts of $v$ and $h$ are the same and using this in the action we obtain that $I_2^{2th}[h(v),v)]=I^{2th}[v]$. On the other hand, making independent variations respect to $h_{\mu \nu}$, we obtain 
\begin{equation}\label{vW}
\epsilon^{\mu\rho\sigma}\partial_\rho v_\sigma ^{\ \nu} = \frac{1}{m} \epsilon^{\mu\rho\sigma}\partial_\rho W_{\sigma} ^{\ \nu}[h]\ \ .
\end{equation}
Substituting (\ref{vW}) in (\ref{In2}) the third order action of the topologically massive gravity emerges. 

\begin{figure}[ht]
\begin{tikzpicture}[sloped]
  \node (a) at ( 0,0) [bag] {$I_3$};

  \node (b) at ( 2,1) [bag] {$I_2$};
  \node (c) at ( 2,-1) [bag] {$I_3^{2th}$};

  \draw [->] (a) to node [above] {} (b);
  \draw [->] (a) to node [below] {} (c);

  \node (d) at ( 4,2) [bag] {$I_1$};
  \node (e) at ( 4,0) [bag] {$I_2^{2th}$};
  \node (f) at ( 4,-2) [bag] {$I_2^{TMM}$};

  \draw [->] (c) to node [above] {} (f);
  \draw [->] (c) to node [below] {} (e);
  \draw [->] (b) to node [above] {} (e);
  \draw [->] (b) to node [below] {} (d);

  \node (g) at ( 6,3) [bag] {$I_0$};
  \node (h) at ( 6,1)   [bag] {$I^{2th}$};
  \node (i) at ( 6,-1) [bag] {$I^{TMM}$};
  \node (j) at ( 6,-3) [bag] {$I^{4th}$};
\node (k) at ( 8,-2) [bag] {$I^{TMM}_{Quad}$};
\node (l) at ( 10,-2) [bag] {$I^{TMM}_{23}$};
  \draw [->] (d) to node [above] {} (g);
  \draw [->] (d) to node [below] {} (h);
  \draw [->] (e) to node [above] {} (h);
  \draw [->] (e) to node [below] {} (i);
  \draw [->] (f) to node [above] {} (i);
  \draw [->] (k) to node [below] {} (i);
  \draw [->] (k) to node [below] {} (j);
 \draw [->] (l) to node [below] {} (i);
  \draw [->] (l) to node [below] {} (j);
\draw[-latex,color=red]
        (g) .. controls +(right:.2cm) and
                                +(right:3cm) ..
            node[near end,above right,color=black] {}
        (i);
\draw[-latex,color=blue]
        (h) .. controls +(left:.2cm) and
                                +(left:3cm) ..
            node[near end,above right,color=black] {}
        (j);
\draw[-latex,color=green]
        (d) .. controls +(left:.2cm) and
                                +(left:4cm) ..
            node[near end,above right,color=black] {}
        (f);
\draw[-latex,color=green]
        (d) .. controls +(right:.2cm) and
                                +(right:4cm) ..
            node[near end,above right,color=black] {}
        (k);
\end{tikzpicture}
\caption{Connections tree between the self-duals models}
\label{tree}
\end{figure}
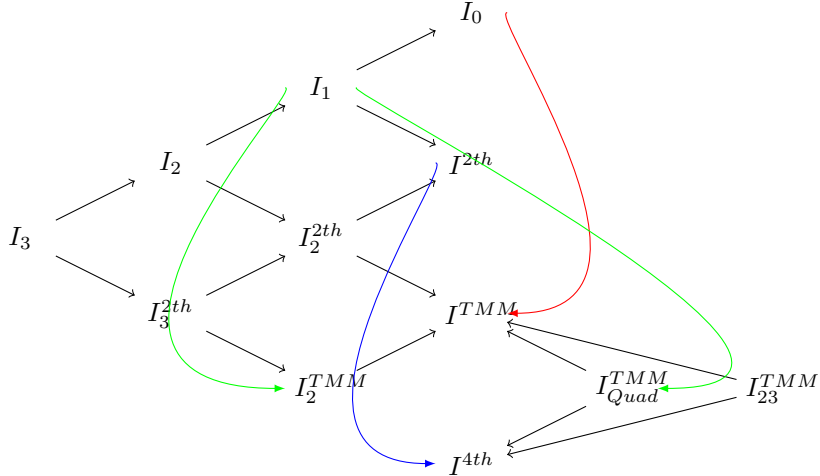

At this point we call the attention of the reader for the first time to Figure (\ref{tree})  where the relations between the different actions are summarized. The curved arrow between $I_0$ and $I^{TMM}$
indicates  the self-dual change of variables mentioned above and the straight arrows show the connections among the canonically equivalent actions.

Before continuing we note that a family $I^{2th}_N$ and in particular $I^{2th}_3$ may also be constructed using the  mechanism of splitting the vector Chern Simons term with the aid of auxiliary field. All these actions are equivalent. The action $I^{2th}_3$ which we write in the form,
\begin{eqnarray}
\label{In3}
I^{2th}_3 [h,v,u]&=& \frac{1}{2m}<\epsilon^{\nu\alpha\beta}\partial_\alpha h_{\beta\rho} \epsilon^{\rho\mu\sigma}\partial_\mu h_{\sigma\nu}>-\frac{1}{4m}<(\epsilon^{\alpha\beta\gamma}\partial_\beta h_{\gamma\alpha})^2>\\
&-&<h_{\mu\lambda} \epsilon^{\mu\nu\rho}\partial_\nu v_\rho^{\ \lambda}>+<v_{\mu\lambda} \epsilon^{\mu\nu\rho}\partial_\nu u_\rho^{\ \lambda}>-\frac{1}{2}<u_{\mu\lambda} \epsilon^{\mu\nu\rho}\partial_\nu u_\rho^{\ \lambda}>\nonumber
\end{eqnarray}
is obtained also directly from $I_3$ by eliminating $w$. If in $I^{2th}_3 [h,v,u]$ we eliminate the field $v$ we obtain a second order master action for $I^{TMM}$ which we denote by $I^{TMM}_2$
\begin{eqnarray}\label{TMM2}
   I_2 ^{TMM}[h,u] &=& -\frac{1}{2m}<h_{\mu\lambda} \epsilon^{\mu\nu\rho}\partial_\nu W_{\rho\lambda}[h]> + \frac{1}{m}<u_{\mu\lambda} \epsilon^{\mu\nu\rho}\partial_\nu W_{\rho\lambda}[h]>\nonumber\\
   &-&\frac{1}{2}<u_{\mu\lambda} \epsilon^{\mu\nu\rho}\partial_\nu u_{\rho\lambda}>.
\end{eqnarray}
Now, making independent variations respect to $u_{\mu\nu}$ we obtain (\ref{vW}) with $u$ instead of $v$, which substituted back in the action allows to recover the action of TMM.  We note that $I^{TMM}_2[h,u]$ may also be obtained making the self dual change of variables $(w\rightarrow w(h),h\longmapsto u)$ of the form (\ref{h(w)}) in $I_1$.  All this relations are reflected in Figure (\ref{tree}).

Contrary to what one could have be expected one cannot arrive from  $I^{TMM}_2$ to a fourth order action and a different scheme should be devised to make contact with the fourth order action recently introduced in Ref.\cite{DalDM2009,AndRBR2010}. This is given by,
\begin{eqnarray}\label{4th}
I^{4th}[h] &=& \frac{1}{2m^3}<\epsilon^{\nu\alpha\beta}\partial_\alpha W_{\beta\rho}(h) \epsilon^{\rho\mu\sigma}\partial_\mu W_{\sigma\nu}(h)>\\&-&\frac{1}{4m^3}<(\epsilon^{\alpha\beta\gamma}\partial_\beta W_{\gamma\alpha(h)})^2>
-\frac{1}{2}<W_{\mu\lambda}(h) \epsilon^{\mu\nu\rho}\partial_\nu W_\rho^{\ \ \lambda}(h)>\nonumber
\end{eqnarray}
Stressing the entangled structure of these systems we note first that this fourth order action is obtained after substituting the self-dual change of variables (\ref{hW}) into (\ref{In}). This is indicated again with the curved arrow which appears between $I^{2th}$ and $I^{4th}$ in Figure (\ref{tree}).

There exist two master actions which connect $I^{TMM}$ and $I^{4th}$. The first is  denoted by $I^{TMM}_{Quad}$ because it use an auxiliary field $u_{\mu\nu}$ with a quadratic Fierz-Pauli like mass term to express the term proportional to the  Einstein action  in  first order. It is given by
\begin{eqnarray}\label{v11}
I^{TMM}_{Quad}[h,u] &=& \frac{m}{2}<u_{\mu\nu} u^{\nu\mu}-u_\mu^\mu u_\nu^\nu> -<u_{\mu\lambda}\epsilon^{\mu\nu\rho}\partial_\nu h_{\rho}^{\lambda}>\nonumber\\&+&\frac{1}{2m^2}<W_{\mu\lambda}[h] \epsilon^{\mu\nu\rho}\partial_\nu W_{\rho}^{\lambda}[h]>,
\end{eqnarray}
It is obviously equivalent to $I^{TMM}$. Making independent variations on $h_{\mu\sigma}$ yields 
\begin{equation}
\epsilon^{\mu\nu\rho}\partial_\nu [u_{\rho\sigma}-\frac{1}{m^2}(W_{\rho\sigma})^{\alpha\beta} W_{\alpha\beta}[h]] = 0.
\end{equation}

Locally, the solution is : $u_{\mu\nu}=\frac{1}{m^2}(W_{\mu\nu})^{\alpha\beta} W_{\alpha\beta}[h] + \partial_\mu \zeta_\nu $. Plugging into the action (after some algebraic manipulations), the fourth order action of Ref.\cite{DalDM2009,AndRBR2010}, is obtained.

The other master action is obtained introducing an auxiliary field $v_{\mu\nu}$ to split the Einstein action without changing its degree. It has the form,

\begin{eqnarray}
I^{TMM}_{23}[h,v] &=& \frac{1}{2} <v_{\mu\nu}\epsilon^{\mu\rho\sigma}\partial_\rho W_{\sigma\nu}[v]>  -  <h_{\mu\nu}\epsilon^{\mu\rho\sigma}\partial_\rho W_{\sigma\nu}[v]>\nonumber\\ &+& \frac{1}{2m}<W_{\mu\lambda}[h] \epsilon^{\mu\nu\rho}\partial_\nu W_{\rho}^{\lambda}[h]>.
\end{eqnarray}
Using the equations of motion we have
\begin{equation}
\epsilon^{\mu\rho\sigma}\partial_\rho W_{\sigma\nu}[v] = \epsilon^{\mu\rho\sigma}\partial_\rho W_{\sigma\nu}[h],
\end{equation}
which establishes the equivalence with $I_{TMM}$.

On the other hand, independent variations on $h_{\mu\nu}$ lead to
\begin{equation}
\epsilon^{\mu\rho\sigma}\partial_\rho [W_{\sigma\nu}[v] -\frac{1}{m}(W_{\sigma\nu})^{\alpha\beta}[W_{\alpha\beta}[h]] = 0\ .
\end{equation}
This has as a solution
\begin{equation}
v_{\mu\nu} = \frac{1}{m}W_{\mu\nu}[h]\ \ .
\end{equation}
Using these results in $I^{TMM}_{23}$ 
the fourth order action is reached.

Here we reach the limit to which the structure discussed in this section may be exploited. If we make the self-dual change of variables (\ref{h(w)}) into the action of the topologically massive gravity (\ref{v}) we obtain a $5th$ order action 
\begin{equation}
I^{5th}[h] = -\frac{1}{2m^2} <W_{\mu\lambda}[h] \epsilon^{\mu\nu\rho}\partial_\nu W_{\rho}^{\lambda}[h]> - \frac{1}{2m^4} <W_{\mu\lambda}[h] \epsilon^{\mu\nu\rho}\Box \partial_\nu W_{\rho}^{\lambda}[h]>,
\end{equation}
which may be written in terms of   symmetric variable $H_{\mu\nu}$ as

\begin{equation}\label{5th}
I^{5th}[H] = -\frac{1}{2m^2} <H_{\mu\nu}C^{\mu\nu}> - \frac{1}{2m^4} <H_{\mu\nu}\Box C^{\mu\nu}>,
\end{equation}
and was shown to  propagate a ghost  in Ref.\cite{BerEHT2010}. 

\section{Duality}

In this section we  make the connection between the master actions
considered in the previous section and duality transformations. 
We use the gauge invariance of some terms in the different models and an auxiliary field, 
which is restricted to be a pure gauge by a constraint equation enforced by a Lagrange multiplier 
to construct dual actions which are found to be the master actions.
We start considering the vector Chern-Simons term in the self dual action (\ref{I0}). This term is invariant under $\delta w_{\mu\nu} =\partial_\mu \zeta_\nu$  We modify the self dual action  with the introduction of a pair of variables $B_{\mu\nu}$ and $ h_{\mu\nu}$) of which $ h_{\mu\nu}$ is a Lagrange multiplier in the form,
\begin{eqnarray}\label{Iodual}
    I^{Dual}_0[w,B,h]&=& \frac{1}{2}<(w_{\mu\lambda} + B_{\mu\lambda})\epsilon^{\mu\nu\rho}\partial_\nu ({w_\rho}^{\lambda}+ {B_\rho}^{\lambda})>\nonumber\\
    &-&\frac{m}{2}<w_{\mu\nu} w^{\nu\mu}-w_\mu^\mu w_\nu^\nu>
    -<h_{\mu\lambda} \epsilon^{\mu\nu\rho}\partial_\nu B_\rho^{\ \lambda}>.
\end{eqnarray}
The field equations which a
\begin{equation}
{\epsilon_\mu}^{\rho\sigma}\partial_\rho[w_{\sigma\nu} +
B_{\sigma\nu}] - m[w_{\mu\nu} - \eta_{\mu\nu}w] = 0,
\end{equation}
\begin{equation}\label{EqRot}
{\epsilon_\mu}^{\rho\sigma}\partial_\rho[w_{\sigma\nu} +
B_{\sigma\nu} - h_{\sigma\nu}] = 0
\end{equation}
and
\begin{equation}\label{constraint}
{\epsilon_\mu}^{\rho\sigma}\partial_\rho B_{\sigma\nu} = 0.
\end{equation}
The constraint (\ref{constraint}) forces $B_{\mu\nu}=\partial_{\mu}\zeta_{\nu}$ to be a pure
gauge which can be gauged away to recover the self dual action.  On the other hand to express the action in terms of the Lagrange multiplier field,  we note that Eq. (\ref{EqRot}) has the local solution
\begin{equation}\label{SolB}
B_{\mu\nu} = -w_{\mu\nu} + h_{\mu\nu} + l_{\mu\nu} ,
\end{equation}
with $\epsilon^{\mu\nu\rho}\partial_\nu l_{\rho\sigma} = 0$.
Substituting (\ref{SolB}) into (\ref{Iodual}), we obtain the
master action $I_{1}[w,h]$ (\ref{I1}) which interpolates 
between $I_0$ and  $I^{2th}[h]$ as discussed in the previous section.

In the same way we may construct the master action $I^{2th}_2[h,v]$ which connects the TMM with the IM starting from the latter. We modify the vector Chern Simons term in the second expression of  $I^{2th}[h]$ in Eq.(\ref{In}) to write
\begin{eqnarray}\label{Imdual}
 I^{2th}_{Dual} &=& \frac{1}{2}<h_{\mu\rho}\epsilon^{\mu\alpha\beta}\partial_\alpha W_{\beta\rho}[h] >
- \frac{1}{2}<[h_{\mu\sigma} +
f_{\mu\sigma}]\epsilon^{\mu\nu\rho}\partial_\nu [h_{\rho\sigma} +
f_{\rho\sigma}]> \nonumber\\ &+& <v_{\mu\lambda}
\epsilon^{\mu\nu\rho}\partial_\nu f_{\rho}^{\lambda}>
\end{eqnarray}
Here, $v_{\mu\nu}$ is the multiplier which enforces the constraint
\begin{equation}
{\epsilon_\mu}^{\rho\sigma}\partial_\rho f_{\sigma\nu} = 0,
\end{equation}
on the auxiliary field  $f_{\mu\nu}$ insuring that it is a pure gauge and in consequence
(\ref{Imdual}) is locally equivalent to (\ref{In}).  The other equations of motion are in this case
\begin{equation}
{\epsilon_\mu}^{\rho\sigma}\partial_\rho [\frac{1}{\mu}W_{\sigma\nu}[h] -
(h_{\sigma\rho} + f_{\sigma\rho})] = 0,
\end{equation}
and
\begin{equation}\label{Consf}
{\epsilon_\mu}^{\rho\sigma}\partial_\rho [v_{\sigma\nu} -
(h_{\sigma\rho} + f_{\sigma\rho})] = 0,
\end{equation}
Solving  $f_{\mu\nu}$ in (\ref{Consf}) as
\begin{equation}
f_{\mu\nu}= - h_{\mu\nu} + v_{\mu\nu} + l_{\mu\nu}
\end{equation}
with ${\epsilon_\mu}^{\alpha\beta}\partial_\alpha l_{\beta\nu} =
0$ and substituting this solution into (\ref{Imdual}), we reach $I^{2th}_2[h,v]$.

Finally, following the procedure outlined above we can construct  the master action $I^{TMM}_{23}$ which connects the TMM with the fourth order model. We modify $I^{TMM}[v]$ by introducing a  pair of non symmetric fields, $f_{\mu\nu}$ and $h_{\mu\nu}$ in the following way:
\begin{eqnarray}\label{TMMnew}
 I &=& -\frac{1}{2}<[v_{\mu\sigma} + f_{\mu\sigma}]\epsilon^{\mu\nu\rho}\partial_\nu W_{\rho\sigma}[v+f]>
+ \frac{1}{2m}<W_{\mu\sigma}[v]\epsilon^{\mu\nu\rho}\partial_\nu W_{\rho\sigma}[v]> \nonumber\\
&+& <h_{\mu\lambda} \epsilon^{\mu\nu\rho}\partial_\nu
{W_{\rho}}^{\lambda}[f]>.
\end{eqnarray}
In this case, $h_{\mu\nu}$ again plays the role of a
multiplier enforcing the constraint that now takes the form
\begin{equation}
{\epsilon_\mu} ^{\rho\sigma}\partial_\rho
W_{\sigma\nu}[f] = 0.
\end{equation}
If instead, we consider the equation of motion obtained taking variations respect to $f$ we get
\begin{equation}\label{EqEin}
{\epsilon_\mu} ^{\rho\sigma}\partial_\rho
W_{\sigma\nu}[v+f-h] = 0.
\end{equation}
 This equation can be solved as
\begin{equation}
f_{\mu\nu}=  h_{\mu\nu} - v_{\mu\nu} + l_{\mu\nu},
\end{equation}
where $l_{\mu\nu}$ satisfies ${\epsilon_\mu} ^{\rho\sigma}\partial_\rho
W_{\sigma\nu [l]} = 0$ and substituting into (\ref{TMMnew}) we reach $I^{TMM}_{23}$

It is interesting to note that it is also possible to construct $I^{TMM}_{23}$ out of $I^{2th}$ in one step
by modifying the first term in (\ref{In}) instead of the second with the auxiliary field restricted by a constraint of the form (\ref{Consf}). The duality transformations considered in this section may also be formulated in the path integral approach in which case the elimination of one of the fields in favor of the Lagrange multiplier is done by means of a quadratic functional integration \cite{AriasF2011,SteJ1997,FraES1994}

\section{Models with two excitations}

In 3D there exist also more than one model of linearized gravity with two excitations. The conventional Fierz-Pauli action is given by 
\begin{equation}\label{FPG}
I_{FP}[h] = \frac{1}{2} <h_{\mu\nu}\epsilon^{\mu\rho\sigma}\partial_\rho W_{\sigma\nu}[h]>  - \frac{m^2}{2} <h_{\mu\nu}h^{\nu\mu} - h^2 >.
\end{equation}
and the fourth order model 
\begin{eqnarray}\label{Fourth}
I^G_4[h] &=&  -\frac{1}{2}<W_{\mu\nu}[v] W_{[v]}^{\nu\mu}-W_{\mu}^\mu [v] W_{\nu }^\nu [v]>  \\
&+&\frac{1}{2m^2}<W_{\mu\lambda}[v]
\epsilon^{\mu\nu\rho}\partial_\nu W_{\rho\lambda} [W(v)]>\nonumber ,
\end{eqnarray}
which was proposed in \cite{AndRBR2010}. 

In 3D describes the ghost free propagation of two physical excitations with opposite helicities. This can be made explicit by observing that (\ref{FPG}) is equivalent to the first order master action action
\begin{equation}\label{FP1}
I_{FP}^{(1)}[h,w] = -\frac{1}{2} <w_{\mu\nu} w^{\nu\mu}-w^2>+<w_{\mu\lambda} \epsilon^{\mu\nu\rho}\partial_\nu h_\rho^{\ \lambda}>
- \frac{m^2}{2} <h_{\mu\nu}h^{\nu\mu} - h^2 >.
\end{equation}
Making in this action the redefinitions
\begin{equation}
h_{\mu\nu} = \frac{1}{\sqrt{2}}[h_{\mu\nu}^1 + h_{\mu\nu}^2 ], \quad w_{\mu\nu} = \frac{m}{\sqrt{2}}[h_{\mu\nu}^1 - h_{\mu\nu}^2 ].
\end{equation}
one has two self-dual models of opposite helicity. This gives a hint that also unitary models with different masses may be consistently constructed in a covariant way. This is discussed below.

The master  action  which connects both may be obtained applying to $I_{FP}$ the dualizing procedure discussed in the previous section.  We modify the action in (\ref{FPG}) in the form,
\begin{eqnarray}
\label{FPGDual}
 I &=& \frac{1}{2}<[h_{\mu\sigma} + f_{\mu\sigma}]\epsilon^{\mu\nu\rho}\partial_\nu W_{\rho\sigma}[h + f]>
- <v_{\mu\sigma}\epsilon^{\mu\nu\rho}\partial_\nu W_{\rho\sigma} [f]> \nonumber\\
&-& \frac{1}{2}<h_{\mu\nu}h^{\nu\mu} - h^2 >.
\end{eqnarray}
Here $v_{\mu\nu}$ is the multiplier which ensures that $f_{\mu\nu}$ is a pure gauge so that (\ref{FPGDual}) is equivalent to $I_{FP}$. The field equation obtained after making independent variations on $f_{\mu\nu}$ is ${\epsilon_\mu} ^{\rho\sigma}\partial_\rho W_{\sigma\nu}[h + f - v] = 0$. The local solution is $f_{\mu\nu} = -h_{\mu\nu} + v_{\mu\nu} + l_{\mu\nu}$ with $l_{\mu\nu}$ satisfying ${\epsilon_\mu} ^{\rho\sigma}\partial_\rho W_{\sigma\nu}[l] = 0$. Substituting this solution we obtain,
\begin{eqnarray}
\label{MFP2}
I_{FP}^{(2)}[h,v] &=& -\frac{1}{2} <v_{\mu\nu}\epsilon^{\mu\rho\sigma}\partial_\rho W_{\sigma\nu}[v]>  +  <h_{\mu\nu}\epsilon^{\mu\rho\sigma}\partial_\rho W_{\sigma\nu}[v]>\\ \nonumber &-& \frac{m^2}{2} <h_{\mu\nu}h^{\nu\mu} - h^2 >.
\end{eqnarray}
Expressed in terms of the symmetric parts of the fields this master action was discussed in \cite{BerEHT2009}. Using the equation of motion,
\begin{equation}
\epsilon^{\mu\rho\sigma}\partial_\rho W_{\sigma\nu}[v] = \epsilon^{\mu\rho\sigma}\partial_\rho W_{\sigma\nu}[h] ,
\end{equation}
to eliminate $v$, we obtain (\ref{FPG}).l
\begin{figure}[ht]
\begin{center}
\begin{tikzpicture}[sloped]
  \node (c) at ( 4,0) [bag] {$I_{FP}$};
  \node (d) at ( 4,-4) [bag] {$I_{FP}^{4th}$};
  \node (e) at ( 6,-2) [bag] {$I_{FP}^{(2)}$};

  \draw [->] (e) to node [above] {} (c);
  \draw [->] (e) to node [below] {} (d);
  
\draw[-latex,color=red]
        (c) .. controls +(left:.2cm) and
                                +(left:3cm) ..
            node[near end,above right,color=black] {}
        (d);
\end{tikzpicture}
\end{center}
\caption{Connections tree between the Fierz-Pauli family of models}
\label{FP2}
\end{figure}
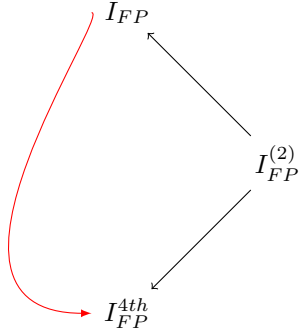

On the other hand, making independent variations on $h_{\mu\nu}$ in the action (\ref{FP2}) leads to
\begin{equation}
h_{\mu\nu} - \eta_{\mu\nu}h =  \frac{1}{m^2}\epsilon_\mu ^{\rho\sigma}\partial_\rho W_{\sigma\nu}[v] .
\end{equation}
which can be solved as
\begin{equation}
h_{\mu\nu} = \frac{1}{m^2}(W_{\mu\nu}^{\rho\sigma})W_{\rho\sigma}[v] = \frac{1}{m^2}S_{\mu\nu}[v].
\end{equation}
where $S_{\mu\nu}$ is the Schouten tensor.
Plugging this expression into (\ref{FP2}), we obtain the fourth order linearized action
The relations between the master actions and the unitary models are shown in Figure 
(\ref{FP2})
The fourth other model may be expressed  in terms only of the symmetric component $H_{\mu\nu}$ in the form,
\begin{equation}\label{FourthSym}
I_{FP}^{4th} = -\frac{1}{2} <H_{\mu\nu}G^{\mu\nu}_{[H]}>  - \frac{1}{2m^2}<[S_{\mu\nu}S^{\mu\nu} - S^2]> \ .
\end{equation} 
The first term is the Einstein action with the ``wrong`` sign and the second is a term with fourth order derivatives ($S_{\mu\nu}S^{\mu\nu} - S^2 = R_{\mu\nu}R^{\mu\nu} - \frac{3}{8}R^2$).

It is worth mentioning that this action can be obtained
directly from the Fierz-Pauli action eq. (\ref{FP2}) by substituting the self dual condition (\ref{h(w)}). This is indicated again with a curved arrow in Figure  (\ref{FP2}).

To discuss the equivalence of the unitary models with two different masses let us recall first the results of reference \cite{AraCK1994}. There it was shown that the model with a Fierz-Pauli mass term and a gravitational Chern-Simons term obtained by spontaneous break of the gauge symmetry of TMM is a non unitary model. On the other hand the model obtained by adding a vector Chern Simons term to the $I_{FP}$ or conversely a Fierz Pauli mass term to IM is unitary with the two excitations having different masses. Its action is given by,

\begin{equation}\label{2m2}
  I^G_2[h] = \frac{1}{2}<h_{\nu\rho}\epsilon^{\nu\alpha\beta}\partial_\alpha W_{\beta\rho}[h] >
-\frac{\mu}{2}<h_{\mu\lambda} \epsilon^{\mu\nu\rho}\partial_\nu h_\rho^{\ \lambda}>-\frac{m^2}{2}<h_{\mu\nu}h^{\nu\mu} - h^2 >.
\end{equation}
with $m$ and $\mu$ two independent mass parameters. The masses of the excitations are,
\begin{equation}
m_ \pm = \frac{1}{2} \mu[\sqrt{1 + 4\frac{m^2}{\mu^2 }} \pm 1 ].
\end{equation}

If we make the substitution of the self-dual condition ($h_{\mu\nu} \rightarrow\frac{1}{m}W_{\mu\nu}[v]$) in (\ref{2m2}) , we reach the so dubbed most general action for spin 2 in three dimensions discussed in \cite{BerEHT2010} whose action is
\begin{eqnarray}\label{I234}
I^G_4[h] &=&  -\frac{1}{2}<W_{\mu\nu}[v] W_{[v]}^{\nu\mu}-w_{\mu}^\mu [v] w_{\nu}^\nu [v]>-\frac{\mu}{2m^2}<W_{\mu\lambda}[v]\epsilon^{\mu\nu\rho}\partial_\nu W_{\rho\lambda}[v]> \nonumber\\
&+&\frac{1}{2m^2}<W_{\mu\lambda}[v]\epsilon^{\mu\nu\rho}\partial_\nu W_{\rho\lambda }[W(v)]> \ ,
\end{eqnarray}
 This action may also expressed  in terms only  of the symmetric component $H_{\mu\nu}(v)$. 
\begin{equation}\label{MG}
I^G_4 [H] = -\frac{1}{2} <H_{\mu\nu}G^{\mu\nu}_{[H]}>  - \frac{1}{2m^2}<[S_{\mu\nu}S^{\mu\nu} - S^2]> + \frac{1}{2\hat{\mu}}<H_{\mu\nu}C^{\mu\nu}_{[H]}>,
\end{equation}
The parameters are related by $\hat{\mu} = \frac{m^2}{\mu}$.

On shell equivalence between (\ref{2m2}) and (\ref{I234}) was discussed in Ref. \cite{BerEHT2010}. To  show the off-shell equivalence of the two systems one could use the (2+1) approach \cite{ArnRDM1962,AraCD1980,AraCS1984} used in Ref. \cite{KhoAS2011} for the vector case or alternatively  construct the master action. In this case to avoid the third order system which was shown to be non-unitary we introduce two new auxiliary field $w_{\mu\nu}$ and $v_{\mu\nu}$ and consider, 
\begin{eqnarray}\label{I2m2}
I^G_M  [h,w,v]&=& -\frac{1}{2}<w_{\mu\nu} w^{\nu\mu}-w_\mu^\mu w_\nu^\nu> + \frac{\mu}{2}<v_{\mu\lambda} \epsilon^{\mu\nu\rho}\partial_\nu v_\rho^{\ \lambda}>\nonumber\\
&+&<w_{\mu\lambda} \epsilon^{\mu\nu\rho}\partial_\nu h_\rho^{\ \lambda}-\mu h_{\mu\lambda} \epsilon^{\mu\nu\rho}\partial_\nu v_\rho^{\lambda}>\\
 &-& \frac{m^2}{2}<h_{\mu\nu}h^{\nu\mu} - h^2 >\nonumber.
\end{eqnarray}
which is clearly equivalent to  (\ref{2m2}). Now it is $h_{\mu\nu}$ which plays the role of a quadratic auxiliary field and its equation of motion allow us to determine that,
\begin{equation}
h_{\mu\nu} = \frac{1}{m^2} W_{\mu\nu} [w - \mu v ].
\end{equation}
Plugging this value of $h_{\mu\nu}$ into (\ref{I2m2}) we reach the following action
\begin{eqnarray}\label{I2m23}
I^G_{M1}[w,v] &=& -\frac{1}{2}<w_{\mu\nu} w^{\nu\mu}-w_\mu^\mu w_\nu^\nu> + \frac{\mu}{2}<v_{\mu\lambda}
\epsilon^{\mu\nu\rho}\partial_\nu v_\rho^{\lambda}>\nonumber\\
&+& \frac{1}{2m^2}<(w_{\mu\lambda}- \mu v_{\mu\lambda}) \epsilon^{\mu\nu\rho}\partial_\nu W_{\rho\lambda} [w - \mu v]>.
\end{eqnarray}

From this action, we have the following field equations
\begin{equation}\label{la}
w_{\nu\mu} - \eta_{\mu\nu} w - \frac{1}{m^2} \epsilon^{\mu\rho\sigma}\partial_\rho W_{\sigma\nu} [w- \mu v]=0
\end{equation}
and
\begin{equation}\label{lb}
\mu \epsilon^{\mu\rho\sigma}\partial_\rho v_{\sigma\nu} - \frac{\mu}{m^2} \epsilon^{\mu\rho\sigma}\partial_\rho W_{\sigma\nu} [w - \mu v]=0.
\end{equation}
which in particular imply,
\begin{equation}\label{al}
w_{\mu\nu} = W_{\mu\nu}[v].
\end{equation}
Using first (\ref{lb}) to obtain 
\begin{eqnarray}\label{I2345}
I^G_{M1}[w,v] &=&  -\frac{1}{2}<w_{\mu\nu} w^{\nu\mu}-w_\mu^\mu w_\nu^\nu> + \frac{1}{2m^2}<w_{\mu\lambda}
\epsilon^{\mu\nu\rho}\partial_\nu W_{\rho\lambda} [w]>\nonumber\\
&-& \frac{1}{2m^2}<w_{\mu\lambda}\epsilon^{\mu\nu\rho}\partial_\nu W_{\rho\lambda}[v]>.
\end{eqnarray}
and then (\ref{al}), we recover the action of the quartic general model (\ref{I234}).

\section{Conclusion}

We presented a unified approach to discuss the equivalence between the known single excitation models of massive gravity in 3D with similar spectrum  using their formulations in terms of non symmetric tensors. We construct a family of master actions which enforce the equivalence between  models of different order in the derivatives  and show how they are obtained using duality transformations. We also discuss  how some of the models are  related more directly by  a self-dual change of variables. 

For the models with two excitations we show that part of the structure remains the same. In particular the master action may also be constructed by a duality transformation. For the models with a mass split, we established the equivalence of the most general fourth order model and the spontaneously broken vector Chern Simons linearized model of massive gravity. 
 
The implications of these relations to the  curved case at the level of actions and solutions remain to be investigated. The generalizations for higher spins of self dual and topologically massive actions, which have been formulated in references \cite{BerEHT2010}, \cite{AraCK21986} and \cite{DarTD1987},  may also be suitable to a similar approach.

\section{Acknowledgments}

This work was partially supported  by DID-USB GID-30 and  PI-03-005753-04 of
CDCH-UCV. JS thanks the members of the  Centro de
F\'{\i}sica Fundamental at ULA for hospitality.


\begin{thebibliography}{References}
\bibitem{DesSJT1982} S.Deser,R.Jackiw and S.Templeton, Ann. Phys. {\bf 140}, 372
(1982).
\bibitem{TowPPV1984} P.K.Townsend,K.Pilch and P.Van Nieuwenhuizen, Phys. Lett. B
{\bf 136}, 38 (1984).
\bibitem{DesSJ1984} S.Deser and R.Jackiw, Phys.Lett.B {\bf 139}, 371 (1984).
\bibitem{AraCK1986} C.Aragone and A.Khoudeir, Phys. Lett. B {\bf 173}, 141 (1986).
\bibitem{AraCK1994} C.Aragone, P.J.Arias and A.Khoudeir, Il Nuov. Cim. B {\bf 108}, 303 (1994).
\bibitem{AraCAK1997} C.Aragone, P.J.Arias and A.Khoudeir, Il Nuov. Cim. B {\bf 112},63 (1997).
\bibitem{DalDM2009} D.Dalmazi and E.L.Mendon\c ca, JHEP {\bf 0909} 011 (2009).
\bibitem{BerEHT2009} E.A.Bergshoeff, O.Hom and P.K.Townsend, Phys. Rev. Lett,
{\bf 102}, 201301 (2009).
\bibitem{BerEHT2009b}  E.A.Bergshoeff, O.Hom and P.K.Townsend, Phys. Rev.,
{\bf D79}, 124042 (2009).
\bibitem{SteK1977} K.Stelle, Phys. Rev. D {\bf 16} 935 (1977).
\bibitem{SteK1977b} K.Stelle, Gen. Rel. and Grav. {\bf 9} 353 (1978).
\bibitem{DesS2009} S. Deser, Phys. Rev. Lett. {\bf 103} 101302 (2009).
\bibitem{BenCM2008} C.M.Bender and P.D.Mannheim, Phys.Rev.Lett. {\bf 100}, 110402
(2008).
\bibitem{AndRBR2010} R. Andriga, E.A.Bergshoeff, M. de Roo, O.Hom, E. Sezgin and P.K.Townsend, Class. Q. Grav.,
{\bf 27}, 025010 (2010).
\bibitem{LSS} W. Li, W. Song and A. Strominger, J. High Energy Phys. 04 082 (2008).
\bibitem{GiaRRS1991} R.Gianvittorio, A.Restuccia and J.Stephany, Mod. Phys. Lett. A  {\bf 6}, 2121 (1991).
\bibitem{RestAS1993} A.Restuccia and J.Stephany, Phys. Lett. B {\bf 305}, 348 (1993).
\bibitem{AriPS1995} P.J.Arias and J.Stephany, J.Math.Phys.  {\bf 36}, 1868 (1995).
\bibitem{DalDM2009b} D. Dalmazi and E.L.Mendon\c ca, Phys. Rev. D {\bf 79} 045025 (2009).
\bibitem{DalD2009} D. Dalmazi, Phys. Rev. D {\bf 79} 085008, (2009).
\bibitem{DalDM2010} D. Dalmazi and E.L.Mendon\c ca, Phys. Rev. D {\bf 82} 105009 (2010).
\bibitem{AriasF2011} P. J. Arias and F. Schaposnik, Int. J. of Mod. Phys. {\bf 26} 2437 (2011).
\bibitem{DalDM2009c} D. Dalmazi and E. L. Mendon\c ca, Phys. Rev. D {\bf 80} 025017 (2009).
\bibitem{AriPR1996} P.J.Arias and A.Restuccia, Phys. Lett. B {\bf 347}, 24 (1995).
\bibitem{AriPRL1996} P.J.Arias, L.Leal and A.Restuccia, Phys. Lett. B {\bf 367}, 170 (1995)
\bibitem{SteJ1997} J.Stephany, Phys. Lett. B {\bf 390}, 128 (1997).
\bibitem{FraES1994} E.Fradkin and F.A.Schaposnik, Phys. Lett. B {\bf 338}, 253,(1994).
\bibitem{ArnRDM1962} R.Arnowit, S.Deser and C.W.Misner in {\it Gravitation:An
introduction to current research},(ed.L.Witten), Wiley NY (1962).
\bibitem{AraCD1980} C.Aragone and S.Deser, Phys. Rev. D {\bf 21} 352, (1980).
\bibitem{AraCS1984} C.Aragone and J.Stephany, Class. Quant. Grav.{\bf 1}, 265
(1984).
\bibitem{KhoAS2011} A.Khoudeir and J.Stephany, Int. J. of Mod. Phys. {\bf 26} 4603 (2011).
\bibitem{BerEHT2010} E.A.Bergshoeff, O.Hom and P.K.Townsend, Ann.of Phys. {\bf 325}, 1118 (2010).
\bibitem{AraCK21986} C.Aragone and A.Khoudeir, Rev. Mex. de F\'{i}s. {\bf 39}, 819 (1993).
\bibitem{DarTD1987} T. Damour and S. Deser, Ann. Inst. Henri Poincar\'{e} {\bf 47} 277 (1987)
\end{thebibliography}
\end{document}